\documentclass[document]{aastex63}
\usepackage{float}
\received{June 1, 2019}
\revised{January 10, 2019}
\accepted{\today}
\submitjournal{AJ}

\shorttitle{22-year cycle magnetic topology}
\shortauthors{Flández et al.}

\begin{document}

\title{A 22-Year Cycle of the Network Topology for Solar Active Regions}

\correspondingauthor{Eduardo Flández}
\email{eduardo.flandez@ug.uchile.cl}

\author[0000-0002-0472-2241]{Eduardo  Flández}
\affiliation{Departamento de Física,\\
Facultad de Ciencias, \\
Universidad de Chile, Santiago, Chile}
\affiliation{Physics Department,\\
	Catholic University of America, Washington, DC, USA}

\author[0000-0002-0211-3151]{Alejandro Zamorano}
\affiliation{Departamento de Física,\\
Facultad de Ciencias, \\
Universidad de Chile, Santiago, Chile}

\author[0000-0003-1746-4875]{Víctor Muñoz}
\affiliation{Departamento de Física,\\
Facultad de Ciencias, \\
Universidad de Chile, Santiago, Chile}



\begin{abstract}

In this paper, solar cycles 21 to 24 were compared using complex network analysis. A network was constructed for these four solar cycles to facilitate the comparison. In these networks, the nodes represent the active regions of the Sun that emit flares, and the connections correspond to the sequence of solar flares over time. This resulted in a directed network with self-connections allowed. The model proposed by Abe and Suzuki for earthquake networks was followed.
The incoming degree for each node was calculated, and the degree distribution was analyzed. It was found that for each solar cycle, the degree distribution follows a power law, indicating that solar flares tend to appear in correlated active zones rather than being evenly distributed.
Additionally, a variation in the characteristic exponent $\gamma$ for each cycle was observed, with higher values in even cycles compared to odd cycles. A more detailed analysis was performed by constructing 11-year networks and shifting them in one-year intervals. This revealed that the characteristic exponent shows a period of approximately 22 years coincident with the Hale cycle,
suggesting that the complex networks provide information about the solar magnetic activity.

\end{abstract}



\section{Introduction} \label{sec:intro}

An active region (AR) is an area of the solar photosphere where the magnetic field strength is higher than in the surrounding photosphere~\citep{howard2011coronal}. These regions have a bipolar structure, with well-ordered flow in two islands of opposite polarity~\citep{priest2014magnetohydrodynamics}. Often, magnetic flux tubes from active regions penetrate the photosphere and emerge as outer loops that re-enter the photosphere, forming sunspots. {The size distribution of active regions follows a power law, indicating that magnetic flux emergence occurs at all size scales on the solar surface~\citep{vlahos2002statistical, daei2017complex, parnell2009power}.}

Solar flares are a sudden, intense, and spatially concentrated release of energy in the corona, resulting in localized heating to temperatures of approximately $10^{7}~\mathrm{K}$ and significant emission of short-wavelength radiation~\citep{charbonneau2001avalanche, howard2011coronal}. These flares often (but not always) occur in active regions and are likely caused by solar magnetic reconnection~\citep{howard2011coronal}. {Among the various proposed mechanisms for the origin of solar flares, magnetic reconnection is widely accepted as the primary cause~\citep{parker1983magnetic, priest2002magnetic, antolin2021reconnection, shen2022origin, garland2022studying, hesse2020magnetic, reeves2022window, csahin2024chromospheric, eriksson2024parker, van2024observations}.}

Solar flares are classified into classes A, B, C, M, or X based on the peak X-ray flux measured by the GOES (Geostationary Operational Environmental Satellite) spacecraft. Each class has a peak flux that is ten times greater than the previous one, with the X class having the highest peak flux at $10^{-4}~\mathrm{W} \mathrm{m}^{-2}$~\citep{priest2014magnetohydrodynamics}. The occurrence of solar flares $(N)$ follows a power law distribution with respect to the total flare energy $W$:
\begin{equation}
	\frac{dN}{dW}\sim W^{-\alpha}\;,
\end{equation}
with $\alpha \sim$ 1.8~\citep{hudson1991solar}.

{The dynamics of solar flares and other phenomena on the sun, make the magnetic field is not as simple in periods of solar maximum as it is in periods of minimum, thus, when it is close to or at solar maximum and the amount of solar flares that occur are greater~\citep{fisk2001behavior}.
Solar flares arise naturally from the interaction between processes in the photosphere, like flux emergence and reconfiguration, and effects in the corona, such as the buildup of electric currents and their resistive dissipation. While initially localized, this redistribution affects nearby areas, setting off a chain reaction that leads to large-scale eruptions of plasma, particles, and magnetic fields~\citep{mcateer2010turbulence}.
It seems that the magnetic field in active regions is the only efficient way to store and release the energy needed to drive solar flares. Since these eruptions are triggered by the stressed magnetic field rooted in the photosphere, studying the magnetic complexity of the photosphere can help predict solar activity and shed light on the underlying physics of the magnetic field~\citep{mcateer2010turbulence}.	
It has been noted that active regions without solar flares generally exhibit a lower level of multifractality compared to those with flares. This indicates that a rise in multifractality signals that a magnetic structure is approaching a critical state~\citep{abramenko2005multifractal}.
Solar flares directly increase the complexity of the evolving magnetic fields in active regions~\citep{aschwanden2006physics}. 
The limited predictability, nonlinearity, and self-organized critical behavior of flares suggest that studying them from a complex systems perspective, such as through complex networks~\citep{gheibi2017solar} or cellular automata based on the Lu-Hamilton model~\citep{lu1991avalanches}, can provide new insights into their underlying complex behavior and event triggering~\citep{lotfi2020ultraviolet}. Consequently, the study of solar physics through complex networks has become a highly active research area~\citep{lotfi2020ultraviolet, taran2022complex, Munoz_EF, gheibi2017solar, zou2014long, najafi2020solar}.}

For example, complex network approaches have been utilized to study the asymmetric activity of solar flares in the two hemispheres~\citep{taran2022complex} and long-term changes in sunspot area asymmetry between the hemispheres~\citep{zou2014long}.
The time series associated with solar flares have also been studied through complex networks~\citep{gheibi2017solar}. These studies explored the behavior of the complex network of solar flares on the solar surface and found that these networks exhibit scale-free characteristics, demonstrating that flares show self-organized criticality.
 Additionally, in Ref.~\citep{Munoz_EF}, a complex network was constructed based on the spatial and temporal evolution of active regions, as determined by image recognition algorithms applied to solar magnetograms taken throughout the entire 23rd solar cycle. 
 
{Therefore, it is valuable to study active regions that produce solar flares using the Abe-Suzuki model~\citep{abe2004scale, abe2006complex}, originally used for studying earthquakes~\citep{Pasten_f, Pasten_h}, to relate these active regions to each other and take advantage of the Sun's complex dynamics.
In the Abe-Suzuki model~\citep{abe2004scale, abe2006complex} each node corresponds to a location, where a seism occurs and a spatial network is formed as time passes. 
A cubic grid of cells with equal size $ \Delta$, covering the geographical area of interest, is used. If a cell contains a seismic event, it is treated as a node in the network, and the connections between nodes represent the temporal sequence of the data. Thus, the first seismic event creates the first node in the network, followed by a link from this node to the second one (the cell where the next seismic event occurs), and so forth~\citep{Pasten_f}. 
In Ref.~\citep{gheibi2017solar}, the surface of the Sun was divided into cells of equal area, with cells containing flares considered as nodes in a network. They discovered that the degree distribution follows a power law, indicating that solar flares form a scale-free network. However, in our study, we focus on hotspot flares confined to well-defined zones, while other flares are measured differently. We also consider a longer time window, examining solar cycles 21 to 24, to study variations between solar cycles.}

This paper is organized as follows. First, in Sec.~\ref{sec:data}, the
datasets used for this study are described. Then, in Sec.~\ref{sec:method}
we described the methods to build complex networks from the data, and
the metrics we use to characterize them. Sec.~\ref{sec:results} presents
the results, and finally, in Sec.~\ref{sec:summary} they are summarized
and discussed.

\section{Data} \label{sec:data}

We consulted all the solar flares that occurred during solar cycles 21 (March 1976 to September 1986), 22 (September 1986 to August 1996), 23 (August 1996 to December 2008), and 24 (January 2008 to December 2019). Among the total number of solar flares available in the Heliophysics Events Knowledgebase (HEK)~\citep{hurlburt2012heliophysics}, only some have an associated active region of origin, while several flares do not originate from active regions but rather from the solar surface itself. These active regions come with an identifier number, which is provided in the HEK database~\citep{hurlburt2012heliophysics}.

\begin{figure}[ht!]
	\plotone{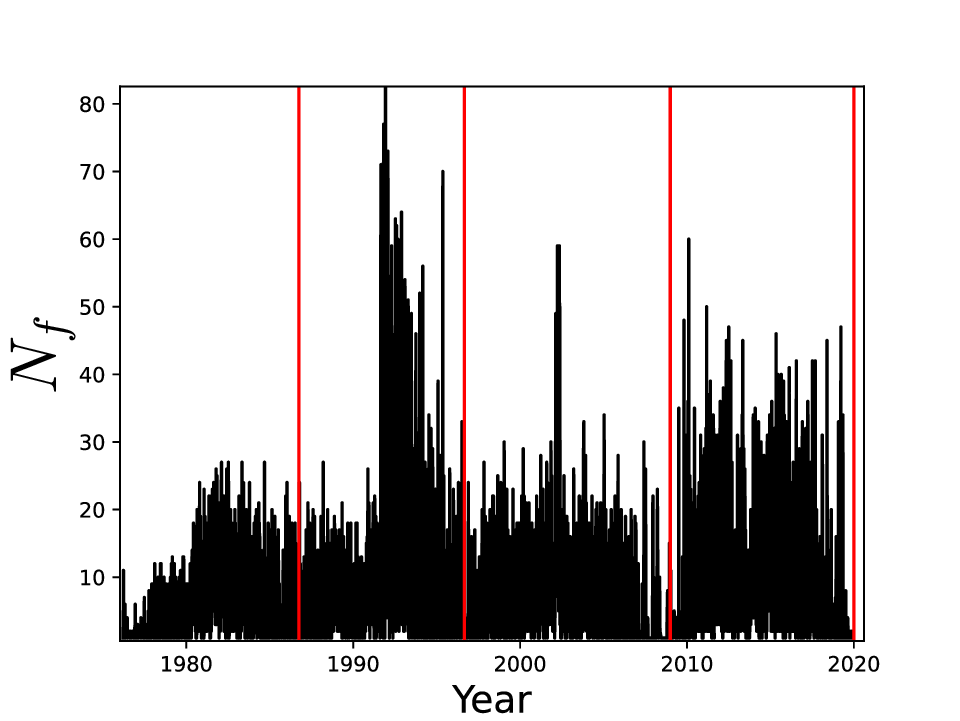}
	\caption{The total number of solar flares $N_f$ emanating from
		active regions, from year 1976 to 2019. Data
		extracted from Space Physics Data Facility,
		NASA/Goddard Space Flight Center~\cite{bornmann1996goes}. The red vertical lines limit solar cycles 21, 22, 23 and 24, respectively.}
	\label{fig:flares_day}
\end{figure}

In Fig.~\ref{fig:flares_day}, we have plotted the number of solar flares per day produced by active regions for solar cycles 21 to 24, showing that cycles 22 and 24 exhibit more solar flare activity than the others. 
In contrast, the number of solar flares that do not originate from active regions for solar cycles 21, 22, 23, and 24 are 7701, 17192, 16152, and 5814, respectively. These figures represent 61\%, 48\%, 38\%, and 80\% of the total solar flares emanating from the entire solar surface. 
In this work, we focus on solar flares originating from active regions, as these can be associated with zones of well-defined magnetic activity. For a study that includes all solar flares, including those outside active regions, please refer to Ref.~\citep{gheibi2017solar}.

To process the data, we use SunPy, an open-source data analysis software for solar physics based on the Python scientific environment~\citep{barnes2020sunpy}. Through this software, we access the list of solar flares available in the HEK database~\citep{hurlburt2012heliophysics}. We also utilize Fido, a unified data search and retrieval tool that provides simultaneous access to various online data sources, including multiple instruments and data products~\citep{barnes2020sunpy}. Specifically, we use the list of flares detected by the GOES X-ray Sensor (XRS) instrument~\citep{bornmann1996goes}.

\section{Method} \label{sec:method}

\begin{figure}[H]
	\centering
	\includegraphics[width=0.5\textwidth]{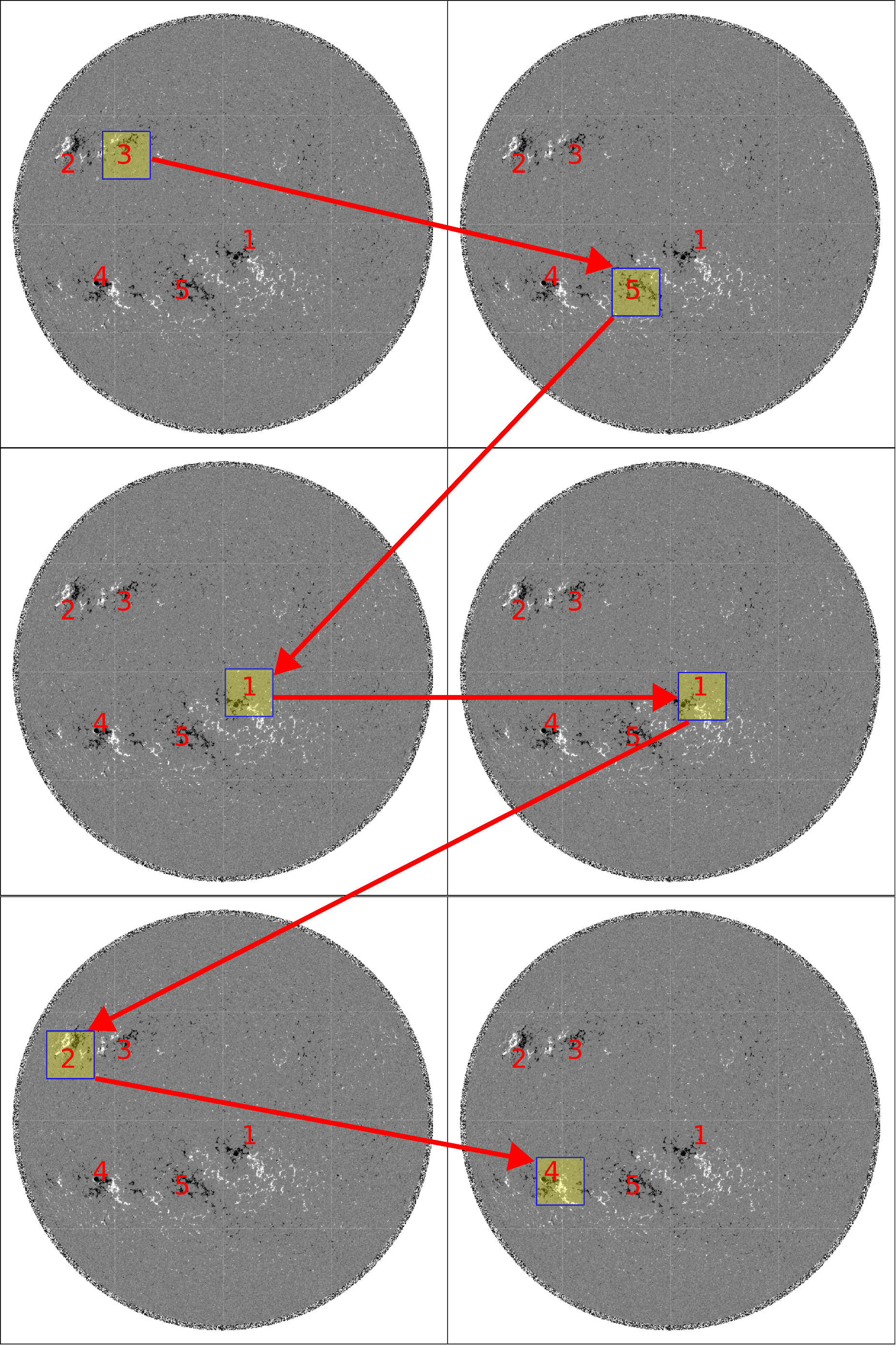} 
	\caption{Illustration of the network construction strategy, using a sequence magnetographs which show active regions associated to solar flares (AR 1--AR 5). Images provided by HMI/SDO mission~\citep{hurlburt2012heliophysics}.}
	\label{fig:metodo}
\end{figure}

To construct the network, we use as nodes only those active regions that produced solar flares. We then connect these nodes based on the sequence in which the solar flares occurred. To illustrate this method, we refer to Fig.~\ref{fig:metodo}. 
For example, if on a given day of the solar cycle, the first solar flare occurs in active region number 3, followed minutes later by a flare from active region number 5, then another from active region number 1, and so on, as depicted in Fig.~\ref{fig:metodo}, the nodes representing active regions AR 1 through AR 5 will be connected in the order of flare occurrence: $3 \rightarrow 5 \rightarrow 1 \rightarrow 1 \rightarrow 2 \rightarrow 4$. 

\begin{figure}[H]
	\plotone{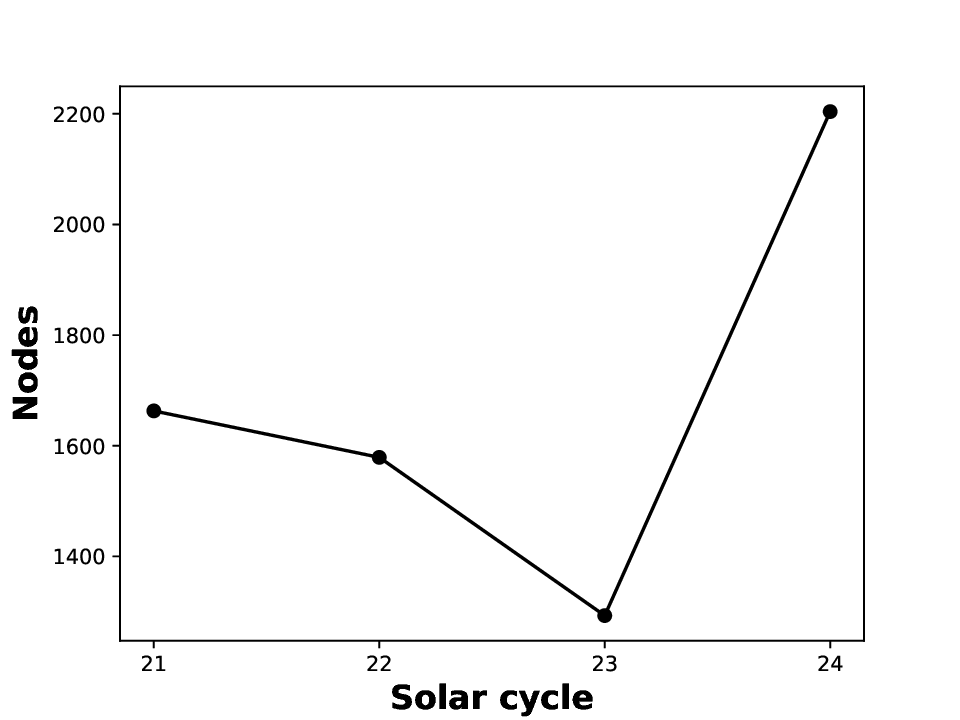}
	\caption{Total nodes for the solar cycles 21 to 24.}
	\label{fig:nodes}
\end{figure}

If two consecutive solar flares originate from the same active region, the node will have a self-connection. This approach, based on the original method proposed by Abe and Suzuki for earthquake networks~\citep{abe2004scale, abe2006complex}, has also been effectively applied to study seismic activity~\citep{Pasten_f, Pasten_h}, solar flares~\citep{gheibi2017solar}, and active regions appearances on the Sun's photosphere~\citep{Munoz_EF}. 
We constructed four {directed} networks, one for each solar cycle: cycle 21 with 1662 nodes, cycle 22 with 1434 nodes, cycle 23 with 1243 nodes, and cycle 24 with 1293 nodes. The number of nodes per solar cycle is shown in Fig.~\ref{fig:nodes}.

In a directed network, each node can be characterized by two types of degrees: the in-degree $k_{i}^{\text{in}}$, which counts the number of incoming connections to the node, and the out-degree $k_{j}^{\text{out}}$, which counts the number of outgoing connections. These degrees can be expressed in terms of the adjacency matrix $A_{ij}$ as follows:
\begin{equation}
	k_{j}^{\text{in}} = \sum_{i}^{n}A_{ij}\;\quad \text{and} \quad k_{i}^{\text{out}} = \sum_{j}^{n}A_{ij}\;.
\end{equation}
First, we construct a directed network and compute both the in-degree $k_{\text{in}}$ and the out-degree $k_{\text{out}}$ for each node. 
We then determine the probability distribution function~$P(k)$, which provides insights into the underlying processes that generate the complex network, in this case, the sequence of flares~\citep{gheibi2017solar,Newman_e,Guillier_b,Barabasi,bianconi2001bose}.

\section{Results}
\label{sec:results}

\begin{figure}[ht!]
	\gridline{\fig{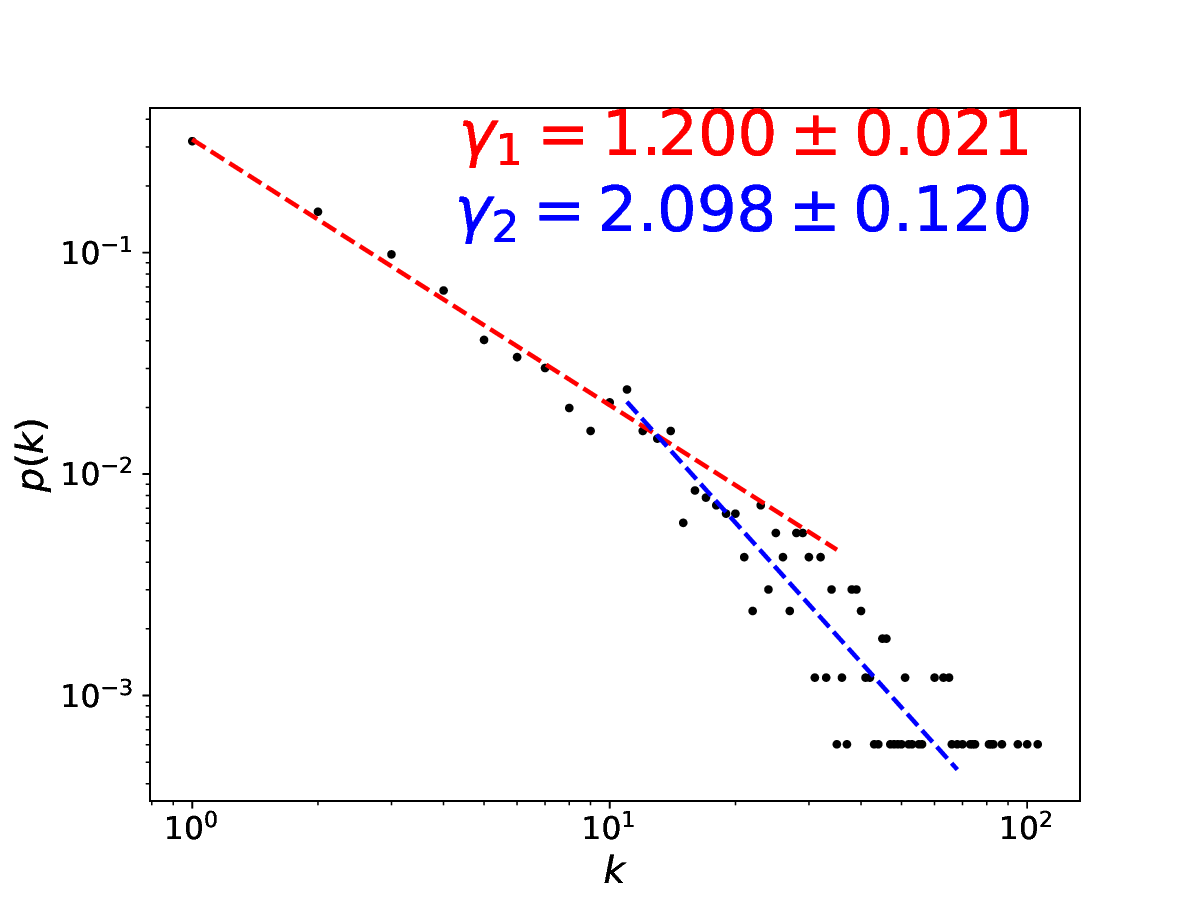}{0.5\textwidth}{(a) SC 21}
		\fig{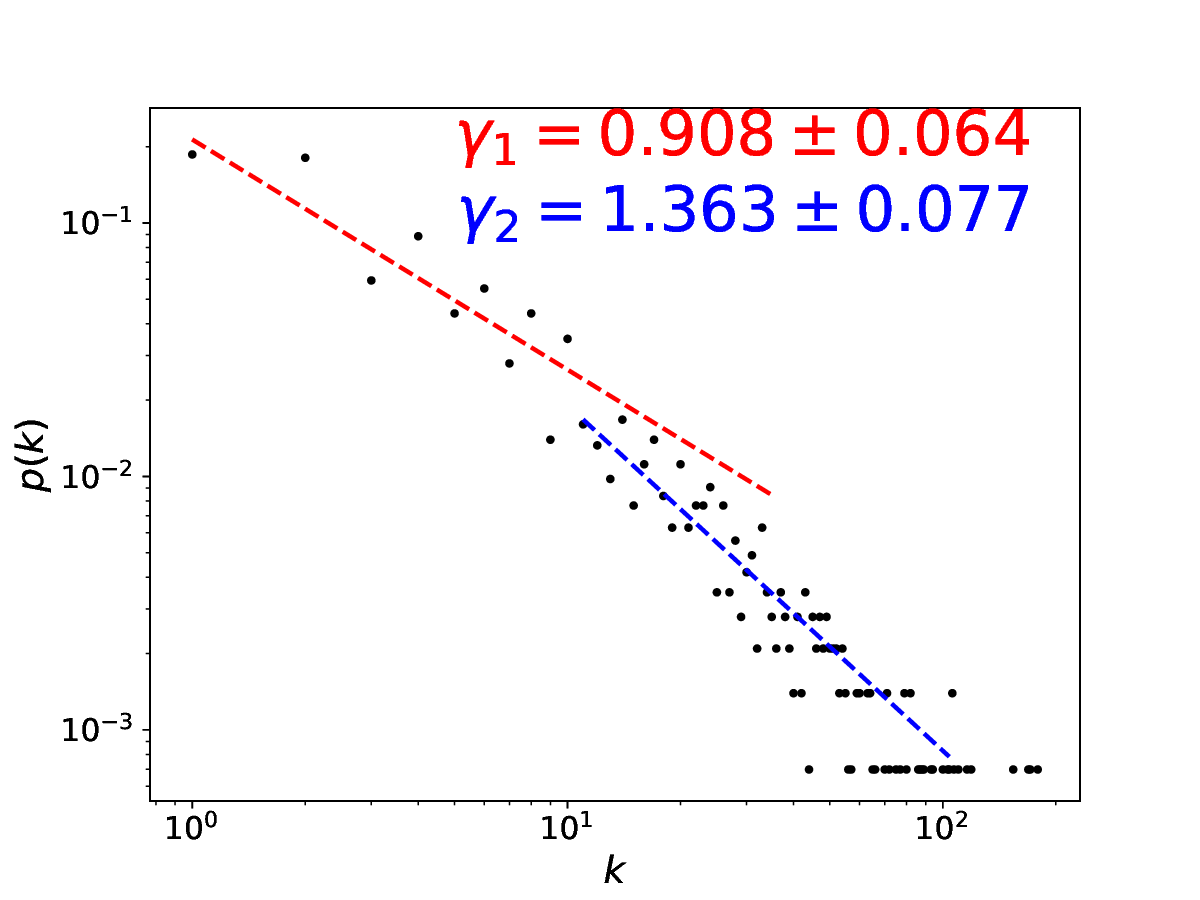}{0.5\textwidth}{(b) SC 22}}
	\gridline{\fig{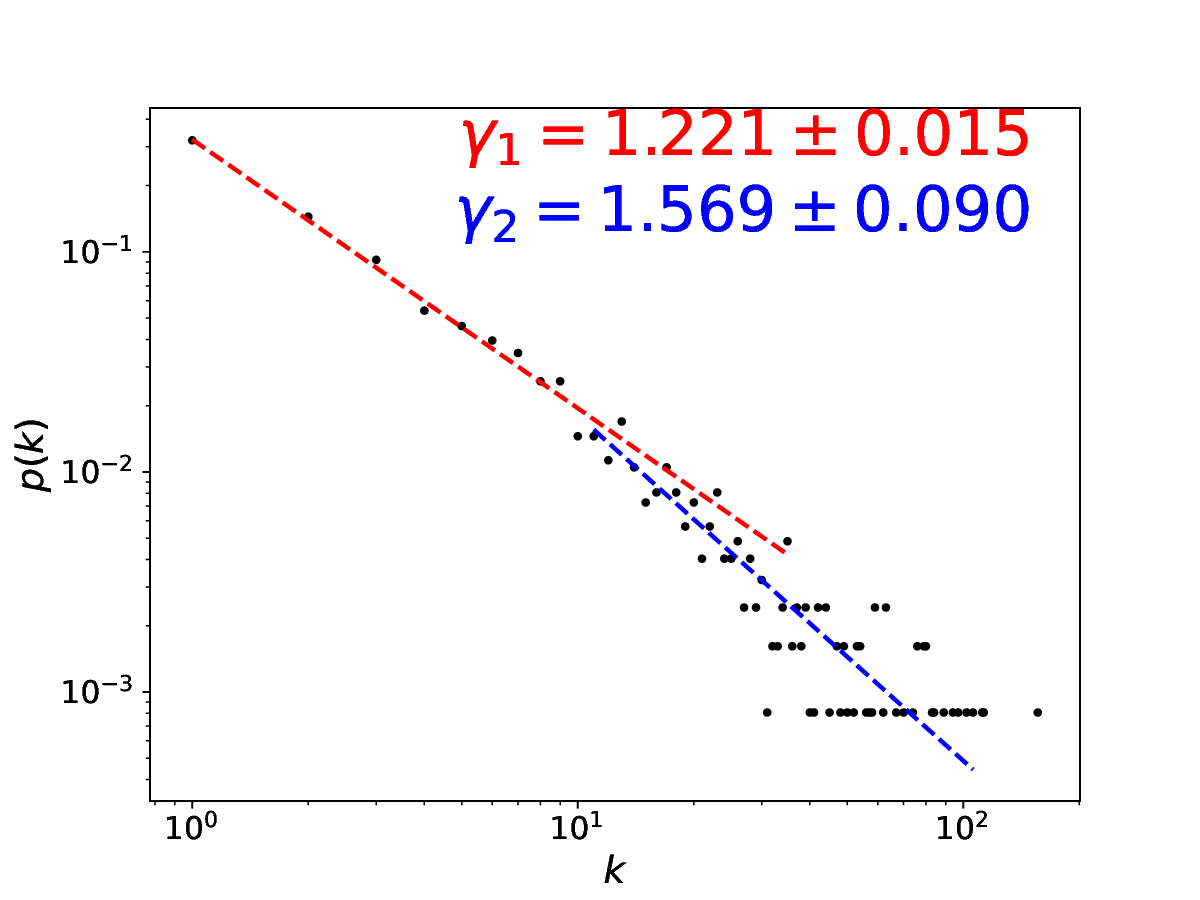}{0.5\textwidth}{(c) SC 23}
		\fig{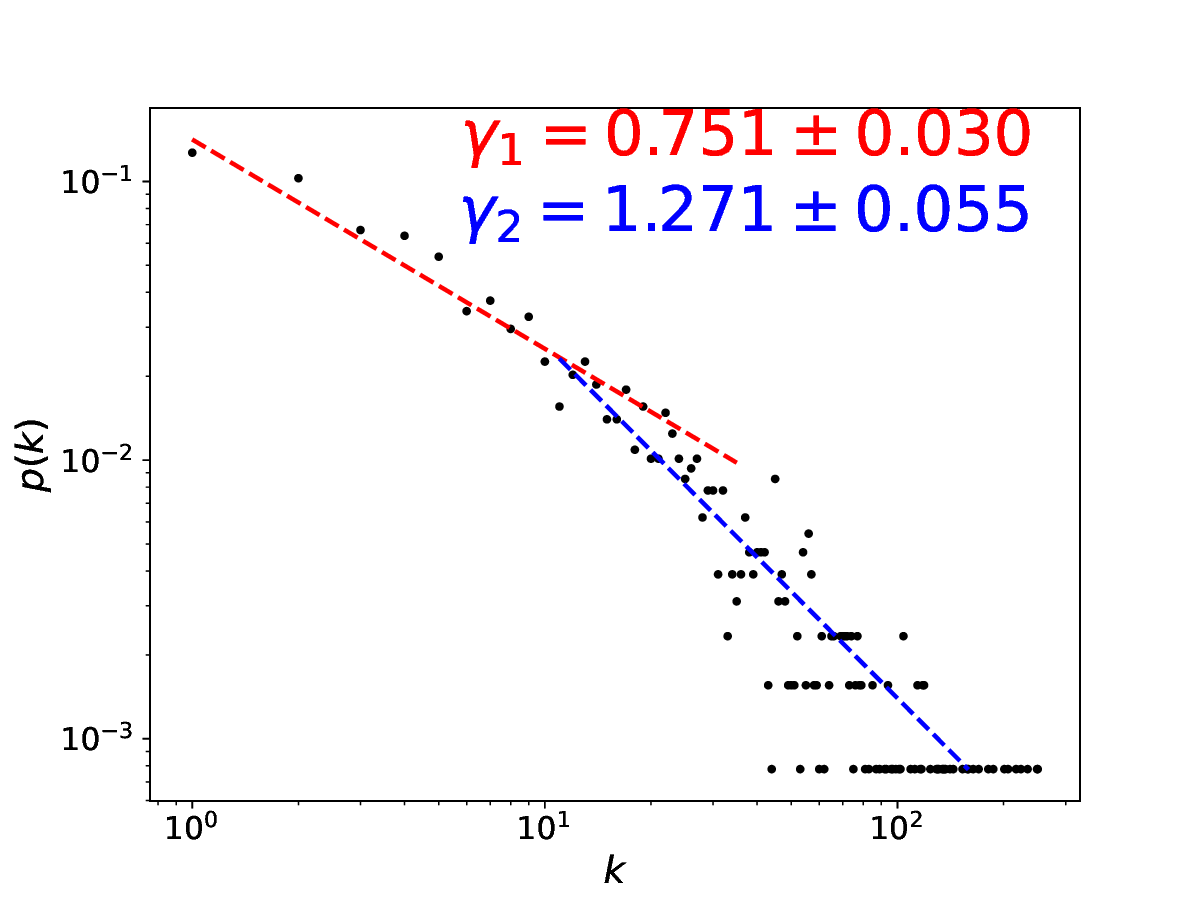}{0.5\textwidth}{(d) SC 24}}
	\caption{{Probability distribution function for the network in-degree corresponding to cycle 21, 22, 23 and 24. Two straight lines can be drawn in the degree distribution, we use red color for the upper one and blue for the lower one.}}
		\label{fig:pdf_in}
\end{figure}

\begin{figure}[ht!]
	\gridline{\fig{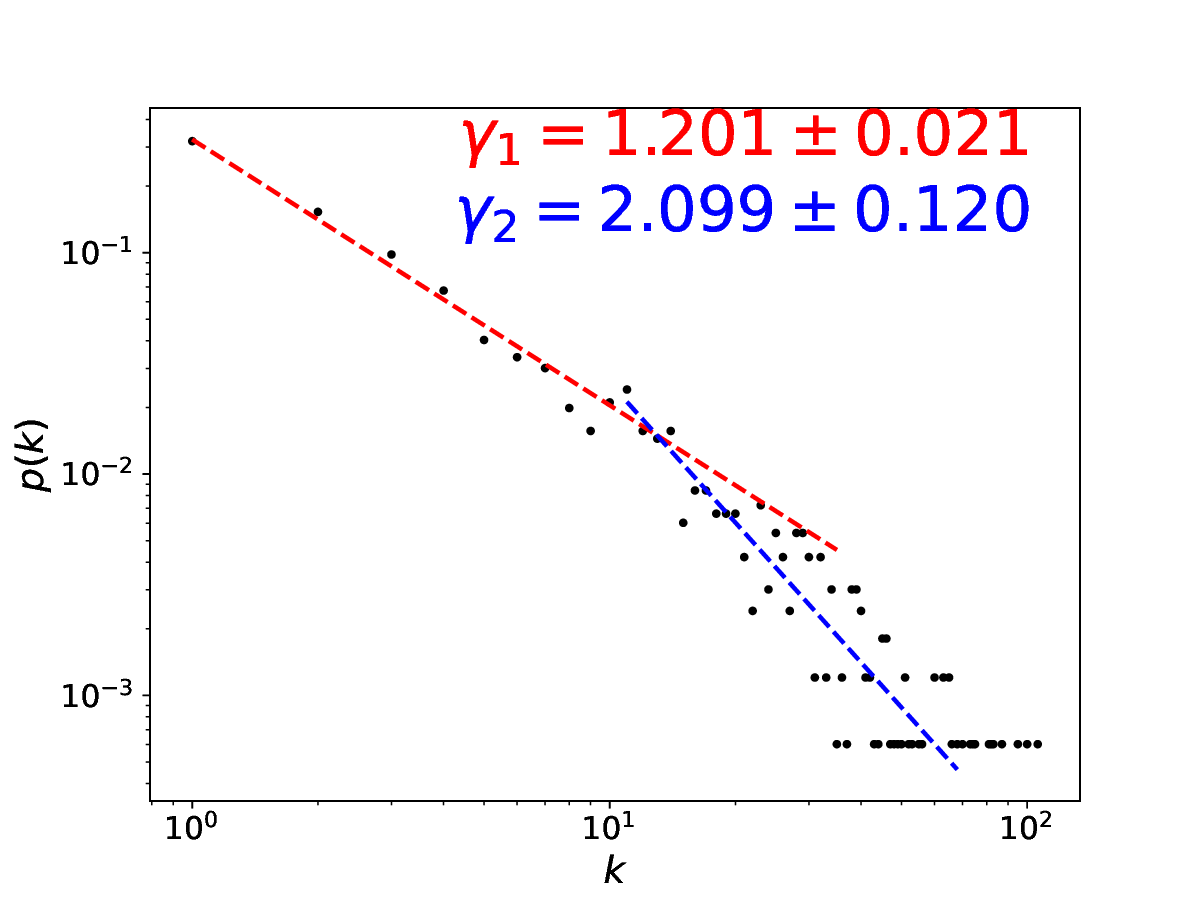}{0.5\textwidth}{(a) SC 21}
		\fig{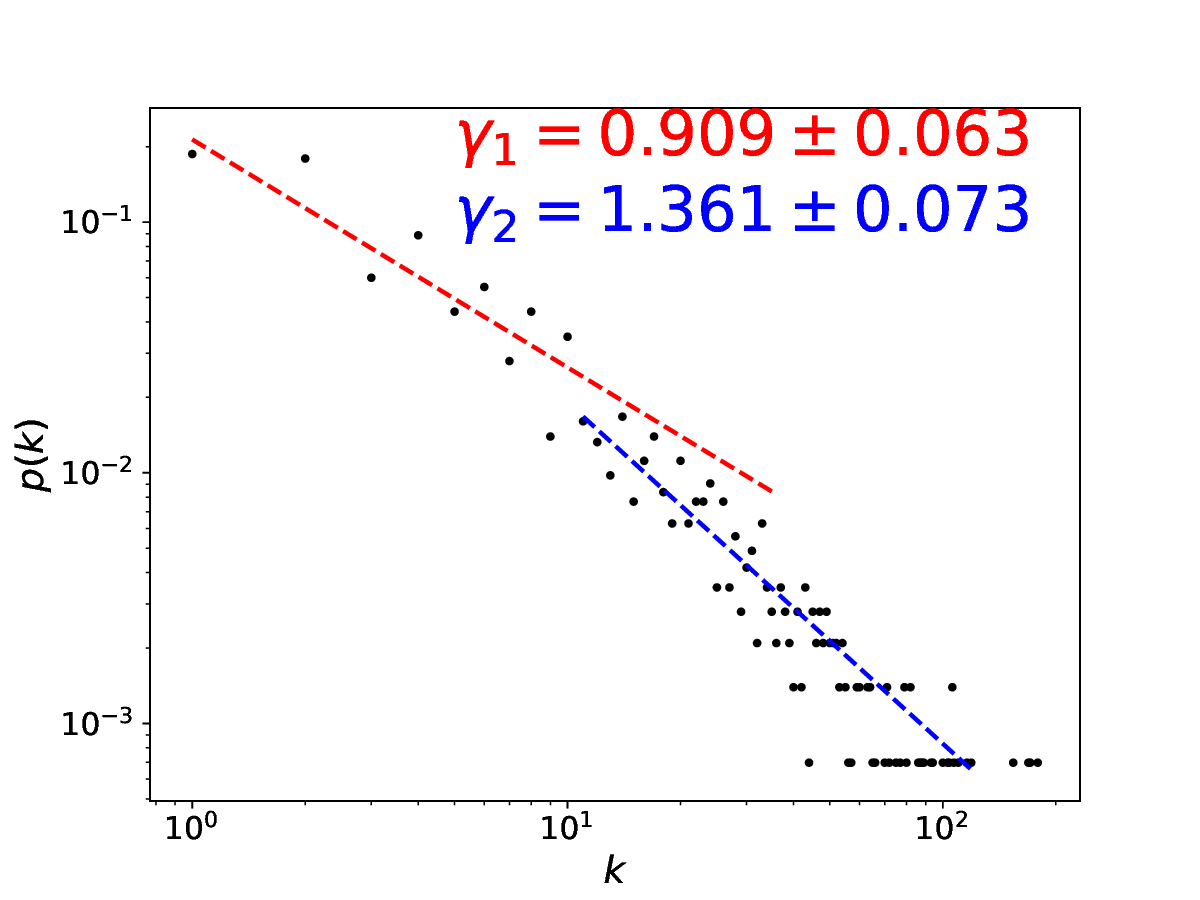}{0.5\textwidth}{(b) SC 22}	}
	\gridline{\fig{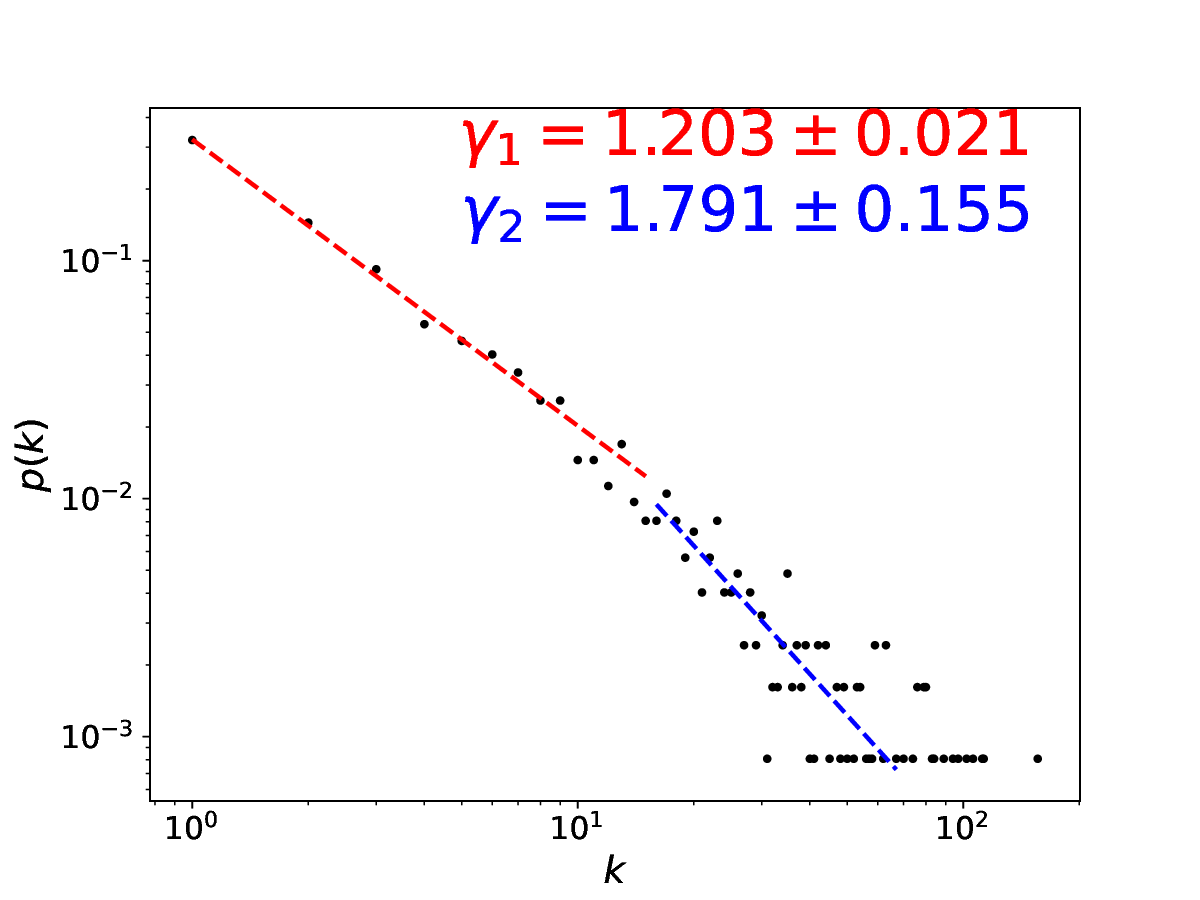}{0.5\textwidth}{(c) SC 23}
		\fig{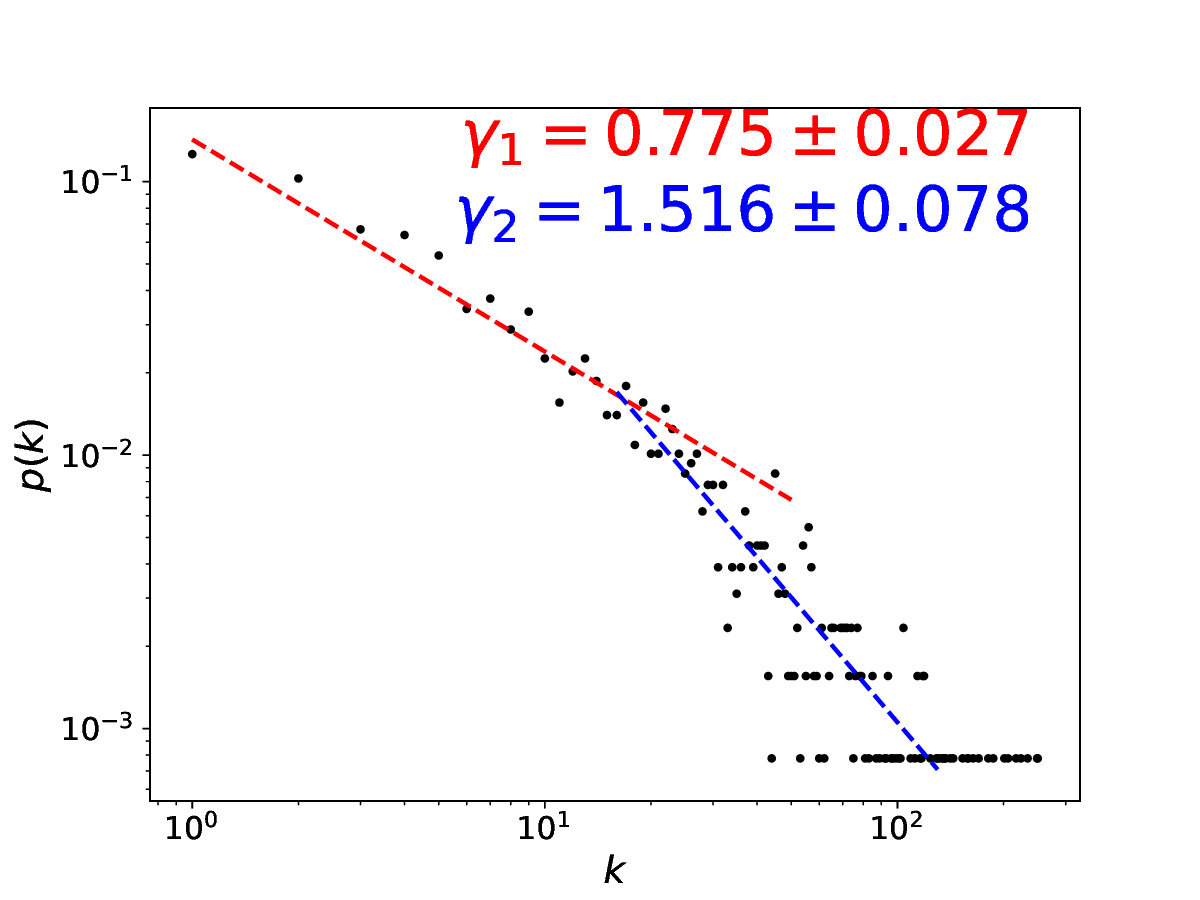}{0.5\textwidth}{(d) SC 24}
	}
	\caption{{Probability distribution function for the network out-degree corresponding to cycle 21, 22, 23 and 24. Two straight lines can be drawn in the degree distribution, we use red color for the upper one and blue for the lower one.}}
	\label{fig:pdf_out}
\end{figure}

Figures~\ref{fig:pdf_in}--\ref{fig:pdf_out} show the plot of the
PDF of degree corresponding to each solar cycle, in logarithmic scale. In all cases, a scale-free behavior can be observed, of the form
\begin{equation} 
	P(k) = ak^{-\gamma} \;,    
\end{equation}
with $a$ a constant, and $\gamma$ is the characteristic
exponent of the network. Figs.~\ref{fig:pdf_in} and \ref{fig:pdf_out}
also show the values of $\gamma$, obtained from linear fits of the
slope for each graph.
This scale-free property obtained is consistent with other works, for
example, with dissipative events such as
earthquakes~\citep{abe2004scale,abe2006complex,Pasten_f,Abe_m,Pasten_h}, solar flares~\citep{gheibi2017solar,najafi2020solar,taran2022complex},
solar proton flux~\citep{mohammadi2021complex} and 
active regions~\citep{daei2017complex}.

\begin{figure}[H]
	\centering
	\includegraphics[height = 0.4\textheight]{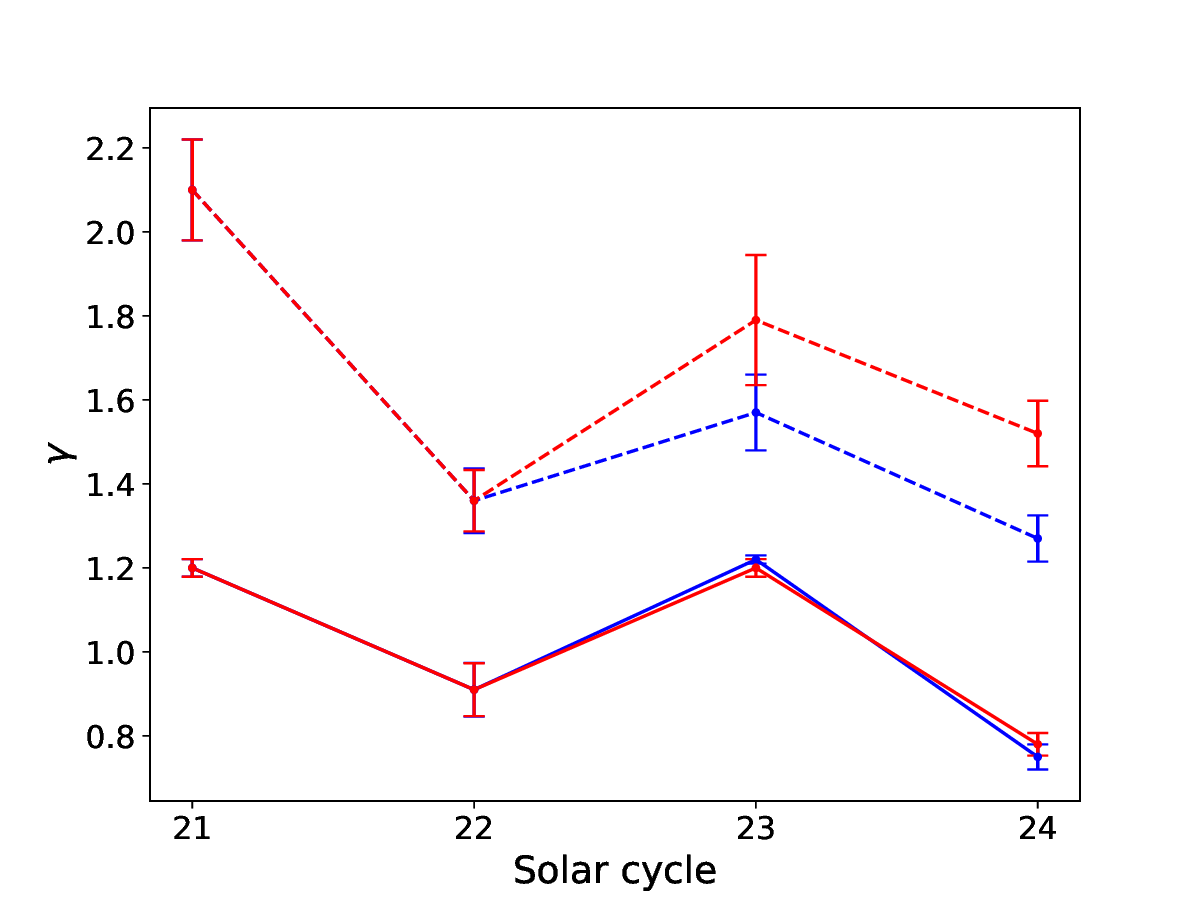}
	\caption{Characteristic decay exponent for each solar cycle. Error
		bar is given by the error in the linear fit as shown in. Blue color indicates in-degree and red color indicates out-degree.
		{The solid lines represent the fit at the upper part of the distribution, while the dashed lines correspond to the fit in the tails. The error bars are larger in the tails compared to the upper part of the distribution.}} 
	\label{fig:gamma}
\end{figure}

We notice that for these four solar cycles, we obtained that odd
cycles have lower $\gamma$ values than even cycles, reminding the
even-odd rule for solar cycles~\citep{Gnevyshev}.  This is highlighted
in Fig.~\ref{fig:gamma}. The differences  
with the previous result in Fig.~\ref{fig:nodes} are interesting. In
Fig.~\ref{fig:nodes}, a very simple 
measure is shown, namely, the number of nodes, which is directly
related to the number of active regions during the cycle.

{On the other hand, an interesting result is found in the characteristic decay exponent for each solar cycle, as shown in Fig.~\ref{fig:gamma}. In the figure, we can observe an alternation of the exponent between even and odd cycles. This indicates that the distribution of connections between nodes is not merely determined by the number of nodes in the network. Otherwise, we would have obtained a connection distribution that closely follows the pattern in Fig.~\ref{fig:nodes} which tells us how many nodes there are, and that regardless of the number of nodes the degree distribution delivers a similar exponent for even and odd cycles. Furthermore, it is not trivial because the distribution could be any type of distribution, be it Poisson, Gaussian, or uniform, which gives other types of information and not the freedom of scale present in this work for the four cycles studied.
If the process leading to flares were completely random, then a
Poisson distribution would be obtained, so the fact that there is a
scale-free distribution is consistent with the underlying process not
being fully random. 
A type of preferential
attachment may be involved~\citep{Barabasi}, such that some active
regions are more likely to yield flares, and highly connected
nodes turn out to be more probable than in a random process, thus
leading to a 
power-law behavior. But, as shown in various
works~\citep{Guillier_b,abe2004scale,abe2006complex,Pasten_f,Abe_m,Pasten_h,gheibi2017solar,najafi2020solar,taran2022complex,mohammadi2021complex,daei2017complex}, the details of the
physical process may be relevant to the actual exponent observed, and
thus the variations seen in Fig.~\ref{fig:gamma} suggest that the
statistical behavior of flares alternates between cycles, not only
regarding the number of active regions, but also regarding the probability of a given region to yield flares.}

\begin{figure}[H]
	\centering
	\includegraphics[width=1.0\linewidth]{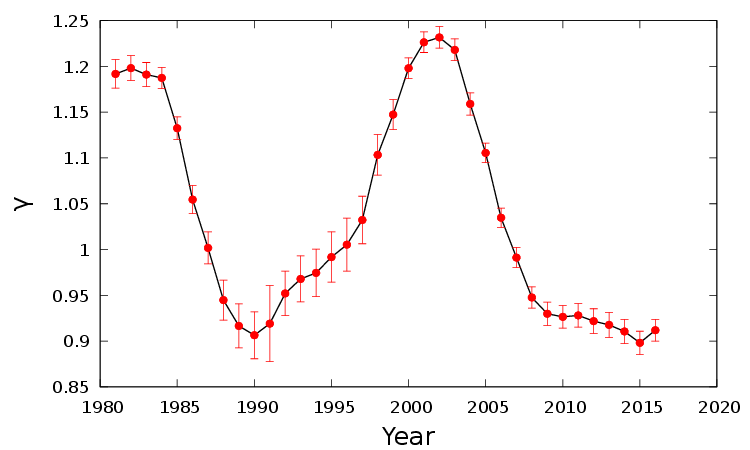}
	\caption{Variation of the characteristic exponent $\gamma$ with error bar for
		sliding windows, each one 11 years in length, and displaced
		by one year. The abscissa for each point correspond to the
		center of each time window.} 
	\label{fig:gammasc}
\end{figure}

A more detailed analysis can be made by means of a sliding window analysis. We take windows spanning 11 years, of the order of the solar cycle duration, shifted one year with respect to the previous one. For instance, the first network was constructed from 1971 to 1982, the second network was constructed from 1972 to 1983, and so on. 
For each window, a scale-free behaviour is found, and the
characteristic exponent is calculated. The result for each window is shown in \ref{fig:gammasc}.

{These distributions correspond to the data for each solar cycle. Since there are approximately 11 years per cycle, we have decided to use data from each cycle to represent the dynamics over these 11 years. On the other hand, if we wanted to extend these values over a longer period (e.g., 110 years, considering 10 cycles), the statistics would improve, as observed in solar flare simulations. In these simulations, as the number of flares approaches 
$10^{6}$, the critical exponent converges to a specific value~\cite{charbonneau2001avalanche}. However, we do not have data on solar flares emitted from active regions. The same reasoning applies to the convolution of the total dataset used to calculate the $\gamma$ time series. These results are robust with respect to the amount of data per cycle.
	}

A notable feature of this plot is that the decay exponent not only oscillates, as seen in Fig.~\ref{fig:gamma}, but also appears to exhibit a period of about two solar cycles. The first and second minima are separated by approximately 22 years, which suggests a connection to the Sun's magnetic cycle known as the Hale cycle. It is widely accepted that this magnetic cycle is driven by the inductive effects of fluid motions within the Sun's interior~\citep{charbonneau2010dynamo}. Hale's law states that the magnetic polarity of active regions alternates between hemispheres and reverses in successive sunspot cycles, resulting in a complete solar magnetic cycle of roughly 22 years. These statistical properties align with the Babcock-Leighton dynamo model~\citep{charbonneau2010dynamo}. The Sun's global magnetic field undergoes a cyclic transition from a global poloidal field to a highly stressed toroidal field during an 11-year cycle. This transition explains the decay of the toroidal field due to differential rotation and its gradual dissipation through meridional flows, ultimately resulting in a relaxed poloidal field at the solar cycle's minimum~\citep{aschwanden2019new}.

After constructing the network of active regions and connecting them according to the temporal sequence of solar flare emissions, and following 11-year moving windows, we have found that the characteristic exponent of the degree distribution varies progressively with a period of 22 years, following the Sun's magnetic activity cycle known as the Hale cycle. This appears only when we study the active regions and flares as a complex system, since neither the occurrence statistics of flares nor the emergence over time of active regions show any structure that could be related to the Hale cycle. This is known in Complex Systems Theory as “emergent properties,” where the whole is more than the sum of its parts.

It would be interesting to study the applications this methodology might have for solar activity studies, whether to complement predictions of solar cycles, solar maxima and minima, or to know when magnetic inversion will occur. Perhaps including parameters such as the characteristic exponent of the degree distribution of the active region networks could help improve models predicting solar activity. For example, it is known that during solar minima, the magnetic field at the poles provides clues about the intensity of the next solar cycle. An interesting work could be to quantify the magnetic field intensities at the poles under some complex network metric, which would allow quantifying how intense the next solar cycle will be, or better yet, if it were possible to have a value for the number of sunspots the next maximum will have. On more local scales, since complex networks provide information on magnetic activity, as we show in this work, it would be interesting to study the complexity of active regions and their relationship with the frequency of solar flare emissions.
This study demonstrate that complex network analysis can provide valuable insights into the evolution of solar activity and reveal universal features that are applicable at any stage of the solar cycle.

\section{Summary}
\label{sec:summary}

{An interesting relationship that can be analyzed from these results from Fig.~\ref{fig:gamma}, is that for even cycles the characteristic exponent is larger. In comparison, for odd cycles, the characteristic exponent is smaller. This suggests that the distribution of active regions emanating flares tends to have preferential zones given an active region emanating a flare for even cycles, also because the figures corresponding to the degree distribution are scale-free. On the other hand, from Fig.~\ref{fig:flares_day}, we can notice that the activity of flares by active regions is higher during even cycles. 
Putting together the two ideas just mentioned, we conclude that flares emanating from active regions are more active during even cycles than during odd cycles and also tend to be distributed in preferential active regions, probably because the magnetic field in these active regions is more complex compared to active regions with fewer events.
This alternation in the solar cycle has been addressed in previous studies. In Ref.~\citep{mursula2023hale}, a systematic alternation of the Hale cycle in the emergence of magnetic flux in both hemispheres is highlighted. It was found that the radio fluxes and sunspot numbers during the peak of odd solar cycles (19, 21, 23) are higher at northern high latitudes compared to southern high latitudes, whereas in even solar cycles (18, 20, 22, 24), the peak fluxes and numbers are greater at southern high latitudes than at northern high latitudes.
This is in agreement with what is observed in Fig.~\ref{fig:gamma}. In addition, this same article describes that the complexity of the magnetic field varies towards the poles depending on which cycle it is in, whether it is even or odd. The alternation of the Hale cycle is also verified for seven solar cycles, for which the last four coincide phenomenologically with those obtained in Fig.~\ref{fig:gammasc}.} On the other hand, from Fig.~\ref{fig:flares_day}, we can notice that the activity of flares by active regions is higher during even cycles. 
Putting together the two ideas just mentioned, we conclude that flares emanating from active regions are more active during even cycles than during odd cycles and also tend to be distributed in preferential active regions, probably because the magnetic field in these active regions is more complex compared to active regions with fewer events.

We find that the characteristic exponent of the degree distribution obeys a power law, which shows an oscillation with a period and a half, with an amplitude of 22 years. This could be a manifestation of the 22-year Hale magnetic cycle.
There is no substantial difference between analyzing in-degree or out-degree in this system,  this together with the out-degree and in-degree networks are correlated, therefore the variation of the critical exponent follows the same pattern during the whole cycle, both the 11-year and the 22-year cycle.\\
{Most importantly, the characteristic exponent of the network degree distribution shows a variation consistent with the Hale cycle, which is a rather interesting result since doing a purely statistical analysis or looking at particular events, we would not have arrived at this result, with this in turn we can appreciate once again that the use of complex systems to study the sun is a useful tool to analyze events that are not evident at first sight from the data.}
 
\acknowledgments

This research was funded by FONDECyT grant number 1242013 (V.M.), and
supported by ANID PhD grant number 21210996 (E.F.) and ANID PhD grant
number 21231335 (A.Z). We are grateful to SDO Data supplied courtesy
of the SDO/HMI consortia. We also thank to
Space Physics Data Facility, NASA/Goddard Space Flight Center.

\bibliography{sample63}{}
\bibliographystyle{aasjournal}

\end{document}